\documentclass[]{article}
\usepackage[affil-it]{authblk}
\usepackage{amsmath}
\usepackage{mathtools}
\usepackage{graphicx}
\usepackage{subcaption}

\begin{document}

\title{Minimal number of discrete velocities for a flow description and internal structural evolution of a shock wave}
\author{Jae Wan Shim}
\affil{Materials and Life Science Research Division, Korea Institute of Science and Technology, and\\Major of Nanomaterials Science and Engineering, KIST Campus, Korea University of Science and Technology, \\5 Hwarang-ro 14-gil, Seongbuk, Seoul 02792, Republic of Korea}
\date{}
\maketitle

\begin{abstract}
A fluid flow is described by fictitious particles hopping on homogeneously distributed nodes with a given finite set of discrete velocities. We emphasize that the existence of a fictitious particle having a discrete velocity among the set in a node is given by a probability. We describe a compressible thermal flow of the level of accuracy of the Navier-Stokes equation by 25 or 33 discrete velocities for two-dimensional space and perform simulations for investigating internal structural evolution of a shock wave. 
\end{abstract}


\section{Introduction}
\label{sec:intro}
It seems intuitively correct to describe fluid flows by using fictitious particles hopping on homogeneously separated nodes with a given finite set of discrete velocities, however, it is not clear how many discrete velocities are needed for the motion of the fictitious particles to satisfy a certain level of accuracy with acceptable stability. This question is clarified by the discrete Boltzmann equation, which is originally developed from the cellular automata to fluid flows. Here we show that we can describe a compressible thermal flow of the level of accuracy of the Navier-Stokes equation by 25 or 33 discrete velocities for two-dimensional space comprised of a square lattice. We look inside the evolution of shock structure by using the fictitious particles.

\section{Rules of collision and movement}
The lattice Boltzmann equation \cite{chen1998lattice, chen2003extended}, originally developed from the cellular automata \cite{rothman2004lattice, shim2010robust} to fluid flows, describes a fluid flow by using the notion of fictitious particles moving their positions and changing their distribution according to a simple rule
\begin{equation}
f_i(x+v_i \Delta t, t+\Delta t)=(1-\omega)f_i(x,t)+\omega f_i^{eq}(x,t)
\end{equation}
where $f_i(x,t)$ is the density of particles having discrete velocities $v_i$ at position $x$ and at time $t$, the reference density distribution $f_i^{eq}(x,t)$ is the density in equilibrium states settled down from $f_i(x,t)$, and $\omega$ adjusts viscosity. Because of the discretized characteristic of the velocity space, $f_i^{eq}(x,t)$ is not the Maxwell-Boltzmann distribution itself but can be expressed by weight coefficients $w_i$ and a polynomial approximated from the Maxwell-Boltzmann distribution. We can recover macroscopic physical properties such as density, velocity, pressure, and temperature from $f_i(x,t)$. To make this particle or lattice-gas method efficient, it is highly desirable to minimize the number of discrete velocities with keeping accuracy and stability.

\begin{figure}
\begin{center}
\includegraphics[scale=0.5]{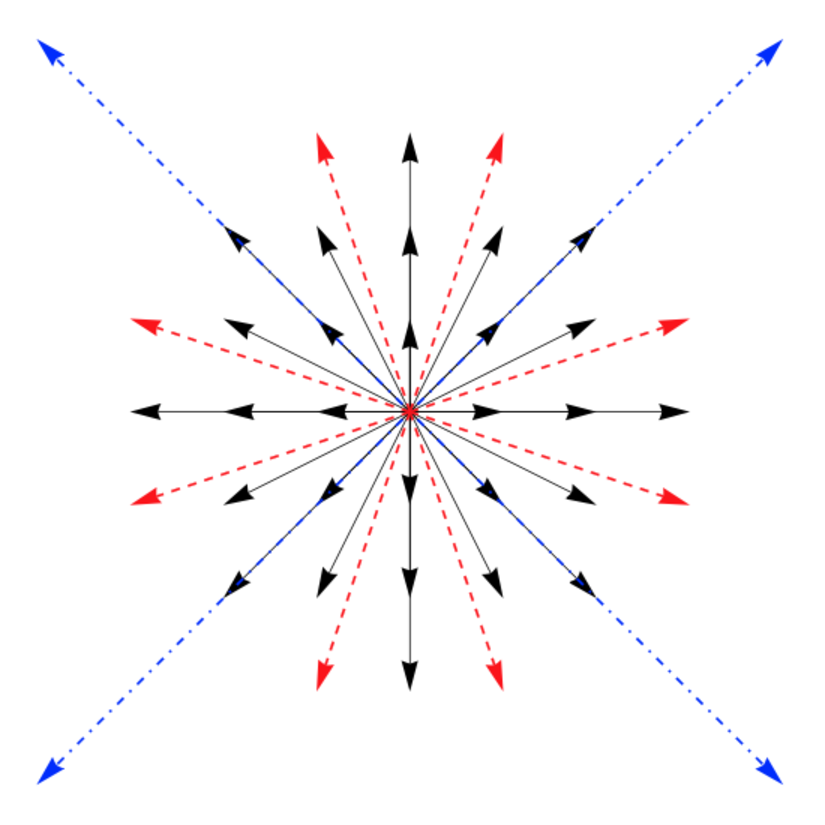}
\end{center}
\caption{The discrete velocities of the 33-, and 37-velocities models are described by the black and the blue (dot-dashed) arrows, and the black and the red (dashed) arrows, respectively. Note that the zero velocity is omitted.}
\label{fig:fig1}
\end{figure}

An important study \cite{philippi2006continuous} showed that compressible thermal flows of the level of accuracy of the Navier-Stokes equation could be recovered by using the lattice Boltzmann equation with 37 discrete velocities in two-dimensional space comprised of a square lattice and this was confirmed again \cite{shan2010general}. However, we can reduce the minimal number by altering discrete velocities. Here, we present a 33-velocities model having the same order of accuracy to the 37-velocities one. As described in Fig.~\ref{fig:fig1}, the vectors of the 33-velocities model are sparsely and widely distributed than those of the 37-velocities one. The discrete velocities of the 33-velocities model $v_i=(v_{i,x},v_{i,y})$ is comprised of $v_1=(0,0)$, $v_2=c(1,0)$, $v_3=c(2,0)$, $v_4=c(3,0)$, $v_5=c(1,1)$, $v_6=c(2,2)$, $v_7=c(4,4)$, $v_8=c(2,1)$ and the other velocities obtained by the symmetry with respect to the $x$-axis, $y$-axis, and $y=x$ where $c=0.819381$, so that the discrete velocities satisfy isotropy. Their corresponding weight coefficients are $w_1\approx 0.161987$, $w_2\approx 0.143204$, $w_3\approx 0.00556112$, $w_4\approx 0.00113254$, $w_5\approx 0.0338840$, $w_6\approx 0.0000844799$, $w_7\approx 3.45552\times 10^{-6}$, $w_8\approx 0.0128169$, and for the other velocities obtained by the symmetry, $w_i=w_j$  if $\|v_i\|=\|v_j\|$. For simplicity, we have presented the approximate values of $c$ and $w_i$ with six significant figures instead of the exact values. Note that this solution can be obtained by the system of equations 
$$\sum_{i=1}^{33}w_i v_{i,x}^a v_{i,y}^b=\Gamma \left(\frac{1+a}{2}\right)\Gamma\left(\frac{1+b}{2}\right)/\pi$$
for $(a,b)=(0,0)$, $(0,2)$, $(2,2)$, $(0,4)$, $(2,4)$, $(0,6)$, $(4,4)$, $(2,6)$, and $(0,8)$ where $\Gamma$ is the Gaussian Gamma function \cite{shim2013multidimensional}. The discretized equilibrium distribution is obtained by the Hermite expansion of the Maxwell-Boltzmann distribution \cite{shim2013obtain} as 
$$f_i^{eq}=\rho w_i \sum_{n=0}^{4}\frac {1}{n!} \bf{a^{(n)}}\cdot \bf{H^{(n)}}$$
where

\begin{align}
\label{HermiteCoefficients}
\bf{a^{(0)}}\cdot \bf{H^{(0)}} &=1, \nonumber \\
\bf{a^{(1)}}\cdot \bf{H^{(1)}} &=2u\cdot v_i,  \nonumber \\
\bf{a^{(2)}}\cdot \bf{H^{(2)}} &=4(u\cdot v_i)^2 + 2(\theta-1)(v_i^2-1)-2u^2,  \nonumber \\
\bf{a^{(3)}}\cdot \bf{H^{(3)}} &=4(u\cdot v_i)\left[ 2(u\cdot v_i)^2 - 3u^2+3(\theta-1)(-2+v_i^2)\right], \nonumber \\
\bf{a^{(4)}}\cdot \bf{H^{(4)}} &=16(u\cdot v_i)^4-48(u\cdot v_i)^2u^2+12u^4 \nonumber \\
&+24(\theta-1)\left[ 2(u\cdot v_i)^2 (v_i^2-3)+(2-v_i^2)u^2\right] \nonumber \\
&+12(\theta-1)^2(v_i^4-4v_i^2+2), \nonumber
\end{align}
$\rho u=\sum v_i f_i$, and $\rho \theta=\sum \|v_i-u_i \|^2 f_i$.

Note that a 25-velocities model for a two-dimensional space is obtained for the level of accuracy of the Navier-Stokes equation by the tensor product of the 5-velocities model as in \cite{shim2017parametric} by using the coefficients of the Lagrange interpolating polynomials expressed by the discrete velocities and using moments of the Maxwell-Boltzmann distribution. The shock tube simulation shows relatively stable and accurate results. This model is less expensive than the 33-velocities model with respect to the computational cost, however, it is less stable in the flow regime of high Mach numbers.

\section{Internal structural evolution of a shock wave}
\label{sec:evolution}

\begin{figure}
\includegraphics[scale=0.8]{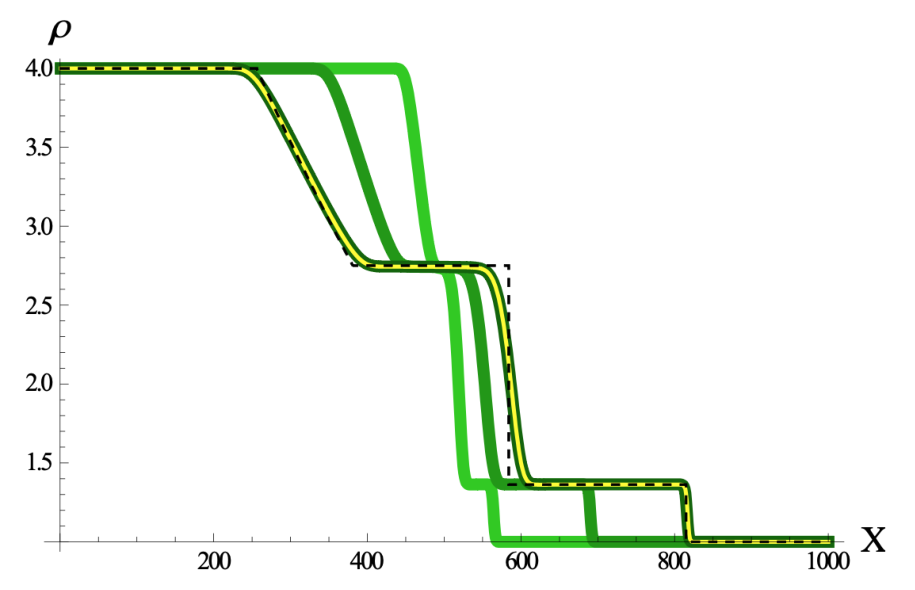}
\caption{Comparison of the scaled density $\rho$ obtained by the 33-velocities model at relative time $t'= 0.2$, $0.6$, and $1$ (from the light green line to the dark), the 37-velocities model at $t'=1$ (yellow), and the analytical solution of the Riemann problem at $t'=1$ (dashed black).}
\label{fig:fig2}
\end{figure}

We illustrate the accuracy and the stability of the 33-velocities model by a shock tube simulation. A two-dimensional shock tube, whose calculation domain is comprised of $1000 \times 8$ nodes, has been simulated by the two models of the 33-velocities and the 37-velocities with the equilibrium distribution $f_i^{eq}$ obtained by the fourth-order Hermite expansion \cite{grad1949note}. Initially, the flow is stationary, and the density and the pressure of the left-half plane are four times higher than those of the right-half plane, while the temperatures are the same in both sides. The left and the right boundary conditions are the same to the left and the right initial conditions, respectively. On the upper and the lower boundaries, the symmetric conditions are used. The constant $\omega$ adjusting viscosity is chosen as $\omega = 1$. The result of the density distribution has no transversal gradient; therefore we show the profile with respect to the longitudinal axis in Fig.~\ref{fig:fig2}. The results obtained by the two models are in excellent agreement. The shock front sharpness of the simulation result is blunt with respect to the analytical solution of the Riemann problem \cite{courant1999supersonic} because of the non-zero viscosity in contrast to the Riemann problem.

\begin{figure}
\includegraphics[scale=0.8]{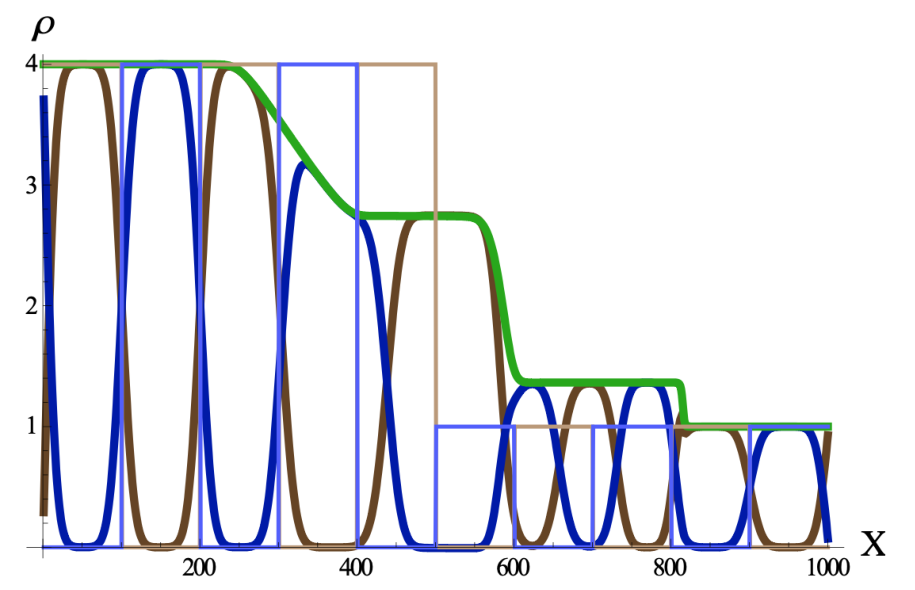}
\caption{Result of the two-component flow simulation. The total density (green) and each component densities (blue and brown) are drawn $t'=1$. Note that the thin blue and brown lines show the initial state and the thick lines show the state after the evolution of time.}
\label{fig:fig3}
\end{figure}

The simple and easy implantation of the multi-component flows is one of the advantages of the flow description by the notion of fictitious particles, on which we just add name tags. We simulate a two-component flow with the previous simulation setup. However, the domain is divided into 10 vertical strips and we fill the two components, alternately. The simulation result is shown in Fig.~\ref{fig:fig3}. The inside structure of the shock is well described.

\section{Multi-component and complex geometry flow simulation}
\label{sec:multi}

\begin{figure}
\includegraphics[scale=0.8]{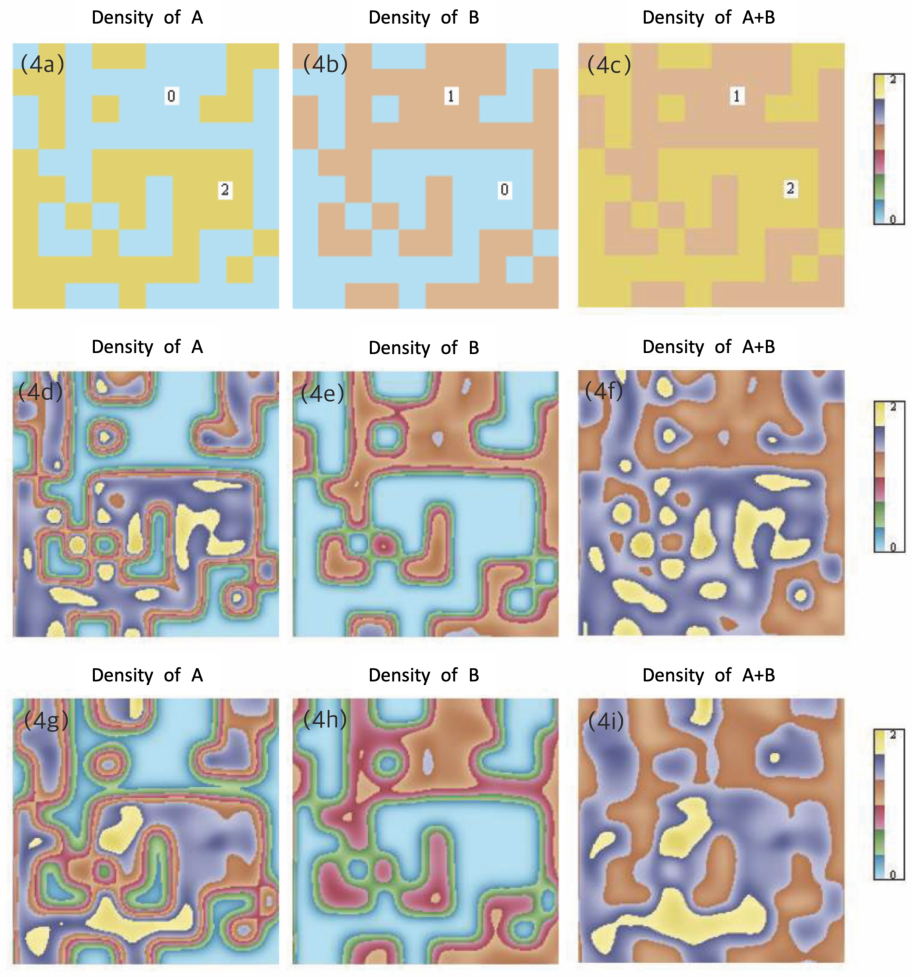}
\caption{Result of the two-component flow having a geometrically complex initial condition. The first, the second, and the third rows show the density distributions at relative time $t'=0$, $0.5$, and $1$, respectively. On the element figures of the first row, the values of density are indicated.}
\label{fig:fig4}
\end{figure}

Another advantage of the notion of fictitious particles is that the simulation method is easily applicable to complex geometry. As an example, the two-component flow is simulated on a plane having a calculation domain comprised of   $200\times 200$ nodes with an arbitrary complex initial condition. The first row of Fig.~\ref{fig:fig4} shows the initial density distribution of the components A and B. On the element figures of the first row, the values of density are indicated. The pressure is the same value to the density. The flow is stationary and the temperature is uniform at the initial moment. 
The simulations illustrate the accuracy and the stability under given conditions. A similar study can be easily done for three-dimensional space.

\section{Conclusion}
\label{sec:conclusion}
We have briefly introduced a method for describing a compressible thermal flow of the level of accuracy of the Navier-Stokes equation by fictitious particles hopping on homogeneously distributed nodes with a given finite set of discrete velocities where the existence of a fictitious particle having a discrete velocity among the set in a node is given by a probability. We have performed simulations for investigating internal structural evolution of shock waves by the method which has advantages in dealing with multi-component flows and complex geometry.

\section*{Acknowledgement}
This work was partially supported by the KIST Institutional Program.

\bibliographystyle{unsrt}
\bibliography{minimalNumber}
\end{document}